\documentclass{wscpaperproc}
\usepackage{latexsym}
\usepackage{graphicx}
\usepackage{mathptmx}
\usepackage[T1]{fontenc}

\usepackage{amsmath}
\usepackage{amsfonts}
\usepackage{amssymb}
\usepackage{amsbsy}
\usepackage{amsthm}
\usepackage{bm}
\usepackage{booktabs}
\usepackage{multirow}
\usepackage{array}

\usepackage[colorlinks=true,urlcolor=blue,citecolor=black,anchorcolor=black,linkcolor=black]{hyperref}

\newtheoremstyle{wsc}{3pt}{3pt}{}{}{\bf}{}{.5em}{}
\theoremstyle{wsc}
\newtheorem{proposition}{Proposition}

\newcommand{\R}{\mathbb{R}}

\begin{document}

\makeatletter
\let\@internalcite\cite
\def\cite{\def\@citeseppen{-1000}\def\@cite##1##2{(##1\if@tempswa , ##2\fi)}\def\citeauthoryear##1##2##3{##1 ##3}\@internalcite}
\def\citeNP{\def\@citeseppen{-1000}\def\@cite##1##2{##1\if@tempswa , ##2\fi}\def\citeauthoryear##1##2##3{##1 ##3}\@internalcite}
\def\citeN{\def\@citeseppen{-1000}\def\@cite##1##2{##1\if@tempswa, ##2)\else{}\fi}\def\citeauthoryear##1##2##3{##1 (##3)}\@citedata}
\def\citeA{\def\@citeseppen{-1000}\def\@cite##1##2{(##1\if@tempswa , ##2\fi)}\def\citeauthoryear##1##2##3{##1}\@internalcite}
\def\citeANP{\def\@citeseppen{-1000}\def\@cite##1##2{##1\if@tempswa , ##2\fi}\def\citeauthoryear##1##2##3{##1}\@internalcite}
\def\shortcite{\def\@citeseppen{-1000}\def\@cite##1##2{(##1\if@tempswa , ##2\fi)}\def\citeauthoryear##1##2##3{##2 ##3}\@internalcite}
\def\shortciteNP{\def\@citeseppen{-1000}\def\@cite##1##2{##1\if@tempswa , ##2\fi}\def\citeauthoryear##1##2##3{##2 ##3}\@internalcite}
\def\shortciteN{\def\@citeseppen{-1000}\def\@cite##1##2{##1\if@tempswa, ##2\else{}\fi}\def\citeauthoryear##1##2##3{##2 (##3)}\@citedata}
\def\shortciteA{\def\@citeseppen{-1000}\def\@cite##1##2{(##1\if@tempswa , ##2\fi)}\def\citeauthoryear##1##2##3{##2}\@internalcite}
\def\shortciteANP{\def\@citeseppen{-1000}\def\@cite##1##2{##1\if@tempswa , ##2\fi}\def\citeauthoryear##1##2##3{##2}\@internalcite}
\def\citeyear{\def\@citeseppen{-1000}\def\@cite##1##2{(##1\if@tempswa , ##2\fi)}\def\citeauthoryear##1##2##3{##3}\@citedata}
\def\citeyearNP{\def\@citeseppen{-1000}\def\@cite##1##2{##1\if@tempswa , ##2\fi}\def\citeauthoryear##1##2##3{##3}\@citedata}
\def\@citedata{\@ifnextchar [{\@tempswatrue\@citedatax}{\@tempswafalse\@citedatax[]}}
\def\@citedatax[#1]#2{\if@filesw\immediate\write\@auxout{\string\citation{#2}}\fi \def\@citea{}\@cite{\@for\@citeb:=#2\do {\@citea\def\@citea{, }\@ifundefined {b@\@citeb}{{\bf ?}\@warning{Citation `\@citeb' on page \thepage \space undefined}}{\csname b@\@citeb\endcsname}}}{#1}}
\def\@citex[#1]#2{\if@filesw\immediate\write\@auxout{\string\citation{#2}}\fi \def\@citea{}\@cite{\@for\@citeb:=#2\do {\@citea\def\@citea{; }\@ifundefined {b@\@citeb}{{\bf ?}\@warning{Citation `\@citeb' on page \thepage \space undefined}}{\csname b@\@citeb\endcsname}}}{#1}}
\def\@biblabel#1{}
\makeatother
\newdimen\bibindent
\bibindent=0.0em
\def\thebibliography#1{\section*{\refname}\list
   {}{\settowidth\labelwidth{[#1]}
   \leftmargin\parindent
   \itemindent -\parindent
   \listparindent \itemindent
   \itemsep 0pt
   \parsep 0pt}
   \def\newblock{}
   \sloppy
   \sfcode`\.=1000\relax}

\WSCpagesetup{Liu, Jiang, and Shen}

\title{SIMULATION-BASED EVALUATION OF ENERGY-CONSTRAINED QUANTUM-CLASSICAL COMPETITION}

\author{\begin{center}Junyu Liu\textsuperscript{1}, Hansheng Jiang\textsuperscript{2}, and Zuo-Jun Max Shen\textsuperscript{3}\\
[11pt]
\textsuperscript{1}Department\ of Computer Science, University of Pittsburgh, Pittsburgh, PA, USA\\
\textsuperscript{2}Rotman School of Management, University of Toronto, Toronto, ON, CANADA\\
\textsuperscript{3}Faculty of Eng.\ and Faculty of Business and Econ., University of Hong Kong, HONG KONG, CHINA\end{center}
}

\maketitle

\vspace{-12pt}
\section*{ABSTRACT}
This paper develops a simulation-based framework for evaluating the energy implications of quantum and classical computing firms competing in a market with limited energy resources. We model providers as differentiated Cournot competitors whose feasible service capacity is induced by technology-specific energy scaling laws: polylogarithmic for quantum algorithms that achieve an equivalent computational target and polynomial for classical emulation. For symmetric groups of quantum and classical firms, the equilibrium reduces to a tractable two-equation system that supports large scenario sweeps over market size, technology mix, and hardware coefficients. We characterize the capacity-constrained Nash equilibrium, prove the existence of a demand scale beyond which quantum service becomes more energy efficient, and report numerical experiments calibrated to trapped-ion and Rydberg platforms. The results identify when quantum energy advantage is only asymptotic and when it becomes operationally relevant.

\section{INTRODUCTION}
Energy costs are becoming increasingly important in the computing industry as large-scale machine learning and simulation workloads are widely deployed. For computing-service providers, energy-efficient operation matters both for market growth and for compliance with sustainability regulations \shortcite{bova2023business}. At the same time, quantum computing, now widely studied in the noisy intermediate-scale quantum (NISQ) era \cite{preskill2018quantum}, is increasingly discussed not only in terms of runtime or complexity-theoretic advantage \cite{SupremacyReview}, but also in terms of its potential energy footprint \cite{auffeves2022quantum}. Moving beyond the conventional notion of quantum supremacy based solely on computational complexity, one may ask whether quantum platforms can deliver comparable algorithmic outputs while consuming substantially less energy.

Several publications and preliminary studies discuss the potential energy benefits of performing equivalent computational tasks on quantum devices (\citeNP{jaschke2022quantum,auffeves2022quantum,meier2023energy}; \shortciteNP{villalonga2020establishing,ikonen2017energy,martin2022energy}). Specific energy cost estimates vary with the type of quantum computing hardware used. For superconducting qubits, the predominant energy expenditure is the cooling required to maintain superconductivity; for instance, Google's quantum devices draw approximately 15\,kW of power during operation, a level largely independent of the number of qubits \shortcite{villalonga2020establishing,arute2019quantum}. To broaden the empirical basis, one can examine platforms whose energy costs scale differently with circuit size, such as Rydberg atom and trapped-ion simulators (\citeNP{jaschke2022quantum}; \shortciteNP{pogorelov2021compact}). These comparisons are important and are complemented by work on the fundamental limitations of noisy quantum optimization \cite{FG20}, but they do not directly answer a central operations question: how do technology-specific energy scaling laws affect market outcomes, resource allocation, and scenario-based performance evaluation?

This paper develops a stylized analytical-simulation framework for that question. We model quantum and classical service providers as differentiated Cournot competitors. Their feasible output is limited by hardware-specific energy laws, so equilibrium quantities, profits, and market shares can be evaluated as functions of the energy budget, market scale, and degree of competition. The framework is simple enough for closed-form analysis and fast enough for large scenario sweeps, which makes it suitable for simulation-based quantum performance evaluation. Using a Cournot competition model constrained by energy usage, we show that quantum computing firms can outperform classical counterparts in profitability and energy efficiency at Nash equilibrium, suggesting that quantum computing may represent a more sustainable pathway for the computing industry. We also find that the energy advantages of quantum computing depend on reaching sufficiently large workload scales.

The paper makes three contributions. First, we formulate a two-group Cournot competition model with technology-specific energy constraints and derive closed-form Nash equilibrium quantities, profits, and market shares for symmetric groups of quantum and classical firms. Second, we characterize the capacity-constrained equilibrium induced by per-firm energy budgets and prove the existence of a crossover demand scale beyond which quantum service consumes less energy than its classical counterpart. Third, we conduct simulation experiments that include a deterministic calibration to trapped-ion and Rydberg hardware parameters and a 40,000-scenario Monte Carlo study, translating the theoretical crossover result into an operational risk profile that quantifies the likelihood and scale at which quantum energy advantage becomes practically relevant.

\section{MODEL}
\subsection{Equivalent-Workload Energy Model}
We interpret $q \geq 2$ as a normalized equivalent-workload scale delivered by a firm over a planning horizon. The index $q$ is not itself the number of physical qubits. Rather, when the calibration is tied to an $n$-qubit logical circuit, we use $n=\log_2 q$, so that $q=2^n$ represents the effective state-space dimension or an equivalent search domain induced by the problem encoding. The model is intentionally stylized: quantum and classical technologies are assumed to achieve comparable algorithmic quality for the same workload index $q$, and the goal is comparative scenario analysis rather than point forecasting.

For a classical platform, the energy required to serve workload $q$ is
\begin{equation}
E_c(q)=\beta_c q^{\alpha},
\label{eq:Ec}
\end{equation}
where $\beta_c>0$ and $\alpha>0$. For a quantum platform, we use
\begin{equation}
E_q(q)=\beta_q \left[\log_2(q)\right]^{\mu},
\label{eq:Eq}
\end{equation}
where $\beta_q>0$ and $\mu>0$. This specification captures the core comparative statics: for comparable algorithmic goals, the classical energy requirement grows polynomially in problem scale while the quantum requirement grows polylogarithmically.

The quantum energy law in \eqref{eq:Eq} should be interpreted as a reduced-form circuit-level scaling model, not as an assumption that a quantum device evaluates all $q$ basis states and reveals all corresponding outputs in one measurement. For a fixed computational target, such as estimating an observable, producing samples from a target distribution, or reaching a prescribed solution-quality tolerance, the parameters $\beta_q$ and $\mu$ absorb the effective scaling of gate count, circuit depth, state preparation, repetitions, measurement shots, and readout overhead with $n=\log_2 q$. Equivalently, one may write $E_q(q)=\beta_q G_q(\log_2 q)$, with the baseline calibration taking $G_q(n)=n^\mu$. The asymptotic comparison below continues to hold whenever the effective quantum overhead satisfies $G_q(\log_2 q)=o(q^\alpha)$, which includes any polynomial overhead in the logical problem size $n$.

We now summarize the physical basis for the baseline calibration. For Rydberg atoms, \citeN{jaschke2022quantum} estimate the energy cost per two-qubit gate at roughly $15\,\mathrm{kJ}$ (based on a gate frequency of $1000\,\mathrm{Hz}$). For an $n$-qubit algorithm using $n^2$ gates in total, with $n=\log_2 q$ (since the workload index $q=2^n$ represents the state-space dimension), the energy is approximately $(\log_2 q)^2\times 15\,\mathrm{kJ}$. For trapped ions, \citeN{jaschke2022quantum} and \shortciteN{pogorelov2021compact} report per-gate energies of about $0.015$-$0.02\,\mathrm{J}$ (for single-qubit and two-qubit gates, respectively, with a M{\o}lmer-S{\o}rensen pulse time of $200\,\mu\mathrm{s}$ and a power draw bounded by $2\times 3.7\,\mathrm{kW}$), giving approximately $(\log_2 q)^2\times 0.0175\,\mathrm{J}$ for the same gate budget.

For the classical side, we consider direct wavefunction simulation using floating-point operations. Using data from the Green500 list of energy-efficient high-performance computing (HPC) clusters \cite{green500}, which report roughly $20\times 10^{9}$ FLOPS per Watt, the classical computational energy for simulating a workload of scale $q$ is approximately $q \times 4\times 10^{-10}\,\mathrm{J}$.

Consolidating these estimates, the baseline calibration uses
\begin{equation}
\alpha=1,\qquad \mu=2,\qquad
\beta_q^{\mathrm{Ion}}=0.0175,\qquad
\beta_q^{\mathrm{Ryd}}=1.5\times 10^{4},\qquad
\beta_c=4\times 10^{-10}.
\label{eq:baseline-beta}
\end{equation}
The quantum coefficients are consistent with the platform-level estimates summarized by \citeN{jaschke2022quantum} and the trapped-ion hardware characteristics reported by \shortciteN{pogorelov2021compact}. The classical coefficient reflects today's energy-efficient HPC consumption \cite{green500}.

\subsection{Competition Model}
Consider $n_q$ quantum firms and $n_c$ classical firms. At a symmetric equilibrium within each technology class, every quantum firm chooses the same output $q_q$ and every classical firm chooses the same output $q_c$. Let $a_q$ and $a_c$ denote net demand intercepts after absorbing non-energy marginal costs into the intercepts. The per-firm profit functions are
\begin{align}
\pi_q(q_q,q_c)
&=q_q\Bigl[a_q-\theta_q q_q-(n_q-1)\gamma_{qq} q_q-n_c\gamma_{qc}q_c\Bigr],
\label{eq:piq}\\
\pi_c(q_q,q_c)
&=q_c\Bigl[a_c-\theta_c q_c-(n_c-1)\gamma_{cc} q_c-n_q\gamma_{cq}q_q\Bigr],
\label{eq:pic}
\end{align}
where $\theta_q,\theta_c>0$ are own-quantity effects and $\gamma_{qq},\gamma_{cc},\gamma_{qc},\gamma_{cq}\geq 0$ measure substitutability. The model allows within-technology and cross-technology interactions to differ. Total quantum market share is
\begin{equation}
s_q=\frac{n_q q_q}{n_q q_q+n_c q_c},
\label{eq:share}
\end{equation}
and the classical share is $1-s_q$.

The following result provides a two-equation representation that avoids a full block-matrix inversion (see Appendix~\ref{app:block} for the derivation).

\begin{proposition}
Define
\begin{equation}
A_q=2\theta_q+(n_q-1)\gamma_{qq},
\qquad
A_c=2\theta_c+(n_c-1)\gamma_{cc},
\qquad
\Delta=A_qA_c-n_qn_c\gamma_{qc}\gamma_{cq}.
\label{eq:ABC}
\end{equation}
If $\Delta>0$ and
\begin{equation}
a_qA_c>a_cn_c\gamma_{qc},
\qquad
 a_cA_q>a_qn_q\gamma_{cq},
\label{eq:positivity}
\end{equation}
then the game has a unique interior Nash equilibrium given by
\begin{align}
q_q^*&=\frac{a_qA_c-a_cn_c\gamma_{qc}}{\Delta},
\label{eq:qqstar}\\
q_c^*&=\frac{a_cA_q-a_qn_q\gamma_{cq}}{\Delta}.
\label{eq:qcstar}
\end{align}
Moreover,
\begin{equation}
\pi_q^*=\theta_q (q_q^*)^2,
\qquad
\pi_c^*=\theta_c (q_c^*)^2.
\label{eq:profitstar}
\end{equation}
\end{proposition}

\begin{proof}
Because each firm's profit is strictly concave in its own decision, the first-order conditions are sufficient. Differentiating \eqref{eq:piq} and \eqref{eq:pic} yields the linear system
\begin{equation}
A_q q_q+n_c\gamma_{qc}q_c=a_q,
\qquad
n_q\gamma_{cq}q_q+A_c q_c=a_c.
\label{eq:foctwo}
\end{equation}
When $\Delta>0$, the system has the unique solution \eqref{eq:qqstar}-\eqref{eq:qcstar}; \eqref{eq:positivity} ensures positivity. Substituting the first-order conditions into \eqref{eq:piq} and \eqref{eq:pic} gives \eqref{eq:profitstar}.
\end{proof}

\section{ENERGY-CONSTRAINED EQUILIBRIUM}
\subsection{Energy Budgets and Capacity Ceilings}
Suppose each quantum firm has an energy budget $B_q$ and each classical firm has an energy budget $B_c$ over the planning horizon. Inverting \eqref{eq:Ec} and \eqref{eq:Eq} gives the technology-specific capacity ceilings
\begin{equation}
\bar q_q(B_q)=2^{(B_q/\beta_q)^{1/\mu}},
\qquad
\bar q_c(B_c)=\left(B_c/\beta_c\right)^{1/\alpha}.
\label{eq:capacity}
\end{equation}
Thus the constrained strategy spaces are $q_q\in[0,\bar q_q(B_q)]$ and $q_c\in[0,\bar q_c(B_c)]$.

Define the truncation operator
\begin{equation}
[x]_0^u = \min\{u,\max\{0,x\}\}.
\label{eq:trunc}
\end{equation}
The best responses are then clipped affine functions.

\begin{proposition}
Under the capacity ceilings \eqref{eq:capacity}, the best responses are
\begin{align}
\mathrm{BR}_q(q_c)&=\left[\frac{a_q-n_c\gamma_{qc}q_c}{A_q}\right]_0^{\bar q_q(B_q)},
\label{eq:brq}\\
\mathrm{BR}_c(q_q)&=\left[\frac{a_c-n_q\gamma_{cq}q_q}{A_c}\right]_0^{\bar q_c(B_c)}.
\label{eq:brc}
\end{align}
If $\Delta>0$, the constrained game has a unique Nash equilibrium.
\end{proposition}

\begin{proof}
Equations \eqref{eq:brq} and \eqref{eq:brc} follow from maximizing the concave profit functions over compact intervals. Define
\begin{equation}
F(q_q)=\mathrm{BR}_q\bigl(\mathrm{BR}_c(q_q)\bigr).
\end{equation}
Ignoring the truncation, the slope magnitude of the composition is
\begin{equation}
\rho=\frac{n_qn_c\gamma_{qc}\gamma_{cq}}{A_qA_c}<1
\end{equation}
because $\Delta>0$. Truncation can only reduce the slope magnitude, so $F$ is a contraction on $[0,\bar q_q(B_q)]$. Hence $F$ has a unique fixed point, and then $q_c=\mathrm{BR}_c(q_q)$ is uniquely determined.
\end{proof}

The constrained equilibrium can be written explicitly by checking which capacity constraints bind. Table~\ref{tab:regimes} summarizes the four regimes.

\begin{table}[tb]
\caption{Piecewise form of the constrained equilibrium.}
\label{tab:regimes}
\centering
\small
\begin{tabular}{>{\raggedright\arraybackslash}p{0.18\linewidth} >{\raggedright\arraybackslash}p{0.50\linewidth} >{\raggedright\arraybackslash}p{0.24\linewidth}}
\toprule
Regime & Equilibrium quantities & Validity check \\
\midrule
No capacity binds & $(q_q^*,q_c^*)$ from \eqref{eq:qqstar}-\eqref{eq:qcstar} & $q_q^*\leq \bar q_q$, $q_c^*\leq \bar q_c$ \\
Quantum binds only & $\left(\bar q_q,\left[\frac{a_c-n_q\gamma_{cq}\bar q_q}{A_c}\right]_0^{\bar q_c}\right)$ & $\bar q_q<\mathrm{BR}_q(q_c)$ \\
Classical binds only & $\left(\left[\frac{a_q-n_c\gamma_{qc}\bar q_c}{A_q}\right]_0^{\bar q_q},\bar q_c\right)$ & $\bar q_c<\mathrm{BR}_c(q_q)$ \\
Both bind & $(\bar q_q,\bar q_c)$ & $\bar q_q<\mathrm{BR}_q(\bar q_c)$ and $\bar q_c<\mathrm{BR}_c(\bar q_q)$ \\
\bottomrule
\end{tabular}
\end{table}

Table~\ref{tab:regimes} is useful in simulation because the equilibrium can be computed without solving a generic complementarity problem. For a given scenario, the analyst evaluates the unconstrained solution, checks whether it violates either energy ceiling, and then moves to the appropriate boundary regime.

\section{ASYMPTOTIC ENERGY ADVANTAGE}
Quantum energy advantage emerges only beyond a sufficiently large computational scale. The next proposition formalizes that statement.

\begin{proposition}
Suppose $a_q=\lambda_q a$ and $a_c=\lambda_c a$ for fixed $\lambda_q,\lambda_c>0$, and assume the interior conditions of Proposition~1 hold for all sufficiently large $a$. Then there exist constants
\begin{equation}
K_q=\frac{\lambda_qA_c-\lambda_cn_c\gamma_{qc}}{\Delta}>0,
\qquad
K_c=\frac{\lambda_cA_q-\lambda_qn_q\gamma_{cq}}{\Delta}>0,
\label{eq:K}
\end{equation}
such that $q_q^*(a)=K_qa$ and $q_c^*(a)=K_ca$. Consequently,
\begin{equation}
\frac{E_q\bigl(q_q^*(a)\bigr)}{E_c\bigl(q_c^*(a)\bigr)}
=
\frac{\beta_q\left[\log_2(K_qa)\right]^{\mu}}{\beta_c(K_ca)^{\alpha}}
\longrightarrow 0
\qquad \text{as } a\to\infty.
\label{eq:ratiozero}
\end{equation}
If the ratio in \eqref{eq:ratiozero} exceeds one at some smaller scale, then at least one crossover scale $a^{\dagger}$ exists at which quantum equilibrium energy first falls below classical equilibrium energy.
\end{proposition}

\begin{proof}
Substituting $a_q=\lambda_q a$ and $a_c=\lambda_c a$ into \eqref{eq:qqstar}-\eqref{eq:qcstar} gives the linear scaling $q_q^*(a)=K_qa$ and $q_c^*(a)=K_ca$. The limit in \eqref{eq:ratiozero} follows because a polylogarithm grows more slowly than any positive power. Existence of a crossover then follows from continuity.
\end{proof}

Proposition~3 is the main reason the energy advantage in this model is scale dependent. Even if the quantum technology has a much larger constant $\beta_q$ at small scales, the asymptotic comparison is eventually dominated by the growth rates.

\section{SIMULATION EXPERIMENTS}
To align the model with the simulation emphasis of the track, we use the closed-form equilibrium as a scenario engine. Each scenario specifies market structure $(n_q,n_c)$, substitution parameters, and hardware coefficients, after which equilibrium quantities and energy use are computed analytically. This makes large sweeps computationally inexpensive.

\subsection{Baseline Deterministic Calibration}
We first consider the symmetric baseline
\begin{equation}
 n_q=n_c=10,
\qquad
\theta_q=\theta_c=2,
\qquad
\gamma_{qq}=\gamma_{cc}=2,
\qquad
\gamma_{qc}=\gamma_{cq}=0.1,
\qquad
 a_q=a_c=a.
\label{eq:baseline-market}
\end{equation}
Under \eqref{eq:baseline-market}, Proposition~1 yields
\begin{equation}
q_q^*(a)=q_c^*(a)=\frac{21}{483}a=\frac{a}{23}\approx 0.04348a.
\label{eq:baselineq}
\end{equation}

\begin{figure}[htb]
\centering
\includegraphics[width=0.867097\linewidth]{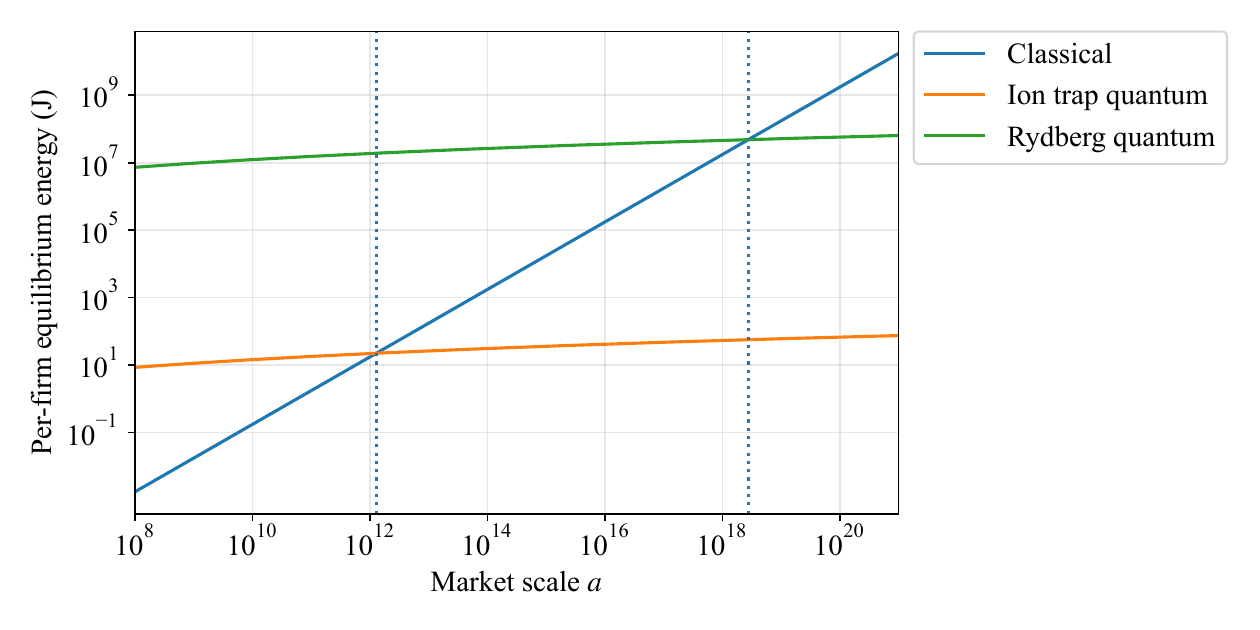}
\caption{Per-firm equilibrium energy under the baseline calibration. The vertical dotted lines indicate the crossover demand scales at which the quantum and classical curves intersect.}
\label{fig:energy-vs-a}
\end{figure}

Figure~\ref{fig:energy-vs-a} reports the resulting per-firm equilibrium energy for classical, trapped-ion, and Rydberg implementations. The crossover scales are approximately $1.28\times 10^{12}$ for ion traps and $2.78\times 10^{18}$ for Rydberg hardware under the baseline calibration.

The figure clarifies two features. First, the eventual quantum advantage is robust in the model because the classical curve is linear in $a$ while the quantum curves grow only logarithmically in $a$. Second, the onset of that advantage depends strongly on the platform-specific constant $\beta_q$. Trapped-ion hardware becomes competitive many orders of magnitude earlier than the Rydberg calibration.

\subsection{Monte Carlo Uncertainty Analysis}
Next, we broaden the deterministic analysis by introducing parameter uncertainty. We generate $40{,}000$ scenarios with the following distributions:
\begin{align*}
&n_q,n_c\sim \mathrm{Unif}\{5,\ldots,20\},
\qquad
\theta_q,\theta_c,\gamma_{qq},\gamma_{cc}\sim U[1,3],\\
&\gamma_{qc}=\gamma_{cq}\sim U[0,0.5],
\qquad
\lambda_q,\lambda_c\sim U[0.8,1.2],
\qquad
\beta_q,\beta_c\text{ perturbed within }\pm 25\%.
\end{align*}
For each scenario, we set $a_q=\lambda_q a$ and $a_c=\lambda_c a$, compute the interior equilibrium, and record the crossover demand scale $a^{\dagger}$ satisfying $E_q(q_q^*(a^{\dagger}))=E_c(q_c^*(a^{\dagger}))$.

Table~\ref{tab:crossover} shows the resulting distribution of crossover scales. The Monte Carlo results preserve the deterministic ordering: trapped-ion quantum computing reaches lower equilibrium energy at a much smaller scale than the Rydberg calibration.

\begin{table}[htb]
\caption{Distribution of the crossover demand scale $a^{\dagger}$ from the Monte Carlo experiment. The baseline column uses \eqref{eq:baseline-market}; the percentile columns come from $40{,}000$ simulated scenarios.}
\label{tab:crossover}
\centering
\begin{tabular}{lcccc}
\toprule
Platform & Baseline & 10th pct. & Median & 90th pct. \\
\midrule
Trapped-ion quantum & $1.28\times 10^{12}$ & $7.16\times 10^{11}$ & $1.56\times 10^{12}$ & $3.23\times 10^{12}$ \\
Rydberg quantum & $2.78\times 10^{18}$ & $1.60\times 10^{18}$ & $3.37\times 10^{18}$ & $6.82\times 10^{18}$ \\
\bottomrule
\end{tabular}
\end{table}

\begin{figure}[htb]
\centering
\includegraphics[width=0.867097\linewidth]{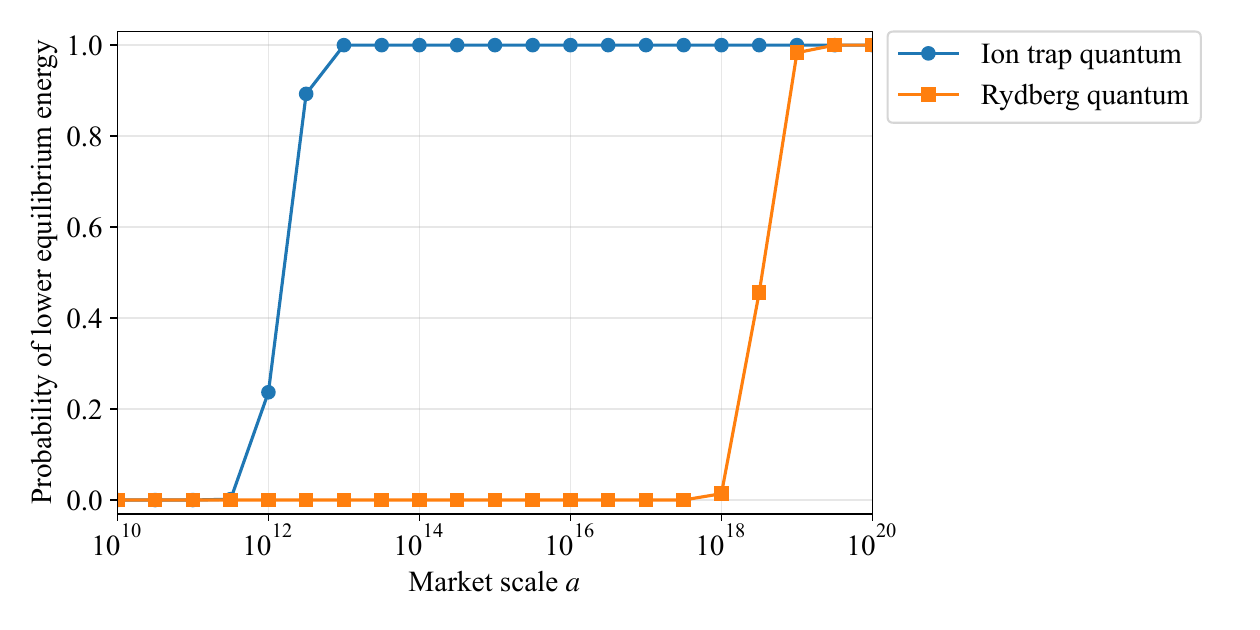}
\caption{Probability that the quantum platform has lower equilibrium energy than the classical platform in the $40{,}000$-scenario Monte Carlo experiment.}
\label{fig:probability}
\end{figure}

Figure~\ref{fig:probability} converts the same experiment into a probability statement. For each market scale $a$, the figure plots the proportion of scenarios in which the quantum platform uses less equilibrium energy than the classical platform. The trapped-ion probability rises rapidly between roughly $10^{12}$ and $10^{13}$, while the Rydberg probability does not rise materially until around $10^{18}$.

From a simulation perspective, Figure~\ref{fig:probability} is useful because it transforms an asymptotic theorem into an operational risk profile. Instead of asking only whether quantum energy advantage exists, one can ask how likely that advantage is under uncertainty in market structure, substitution, and hardware coefficients.

\subsection{Asymmetric-Firm Scenario Check}
To check that the conclusions do not rely entirely on within-group symmetry, we also evaluate a fully asymmetric instance using the general formulation in Appendix~\ref{app:asym}. We retain $N=20$ firms, with ten quantum and ten classical providers, but perturb firm-level demand intercepts, own-quantity effects, and substitution coefficients around the baseline. Specifically, for $i,j=1,\ldots,20$, we set
\begin{equation}
\lambda_i=1+0.02\sin(0.7i),
\qquad
\theta_i=2\left[1+0.02\cos(0.9i)\right],
\label{eq:asym-lambda-theta}
\end{equation}
and, for $i\neq j$,
\begin{equation}
\gamma_{ij}=\begin{cases}
2\left[1+0.01\sin(0.37i+0.53j)\right], & \text{if } i,j \text{ are in the same technology class},\\
0.1\left[1+0.1\cos(0.41i+0.29j)\right], & \text{otherwise}.
\end{cases}
\label{eq:asym-gamma}
\end{equation}
For each market scale $a$, we set $a_i=\lambda_i a$ and compute the interior equilibrium from $q^*(a)=\Gamma^{-1}(\lambda a)$, where $\lambda=(\lambda_1,\ldots,\lambda_N)^\top$.

The matrix $\Gamma$ is nonsingular in this instance, and all equilibrium coefficients $k_i=q_i^*(a)/a$ are positive. Using the baseline $\alpha=1$ and $\mu=2$, the aggregate crossover scale solves
\begin{equation}
\sum_{i\in\mathcal Q}\beta_q[\log_2(k_i a)]^2
=
\sum_{i\in\mathcal C}\beta_c k_i a,
\label{eq:asym-crossover}
\end{equation}
where $\mathcal Q$ and $\mathcal C$ denote the sets of quantum and classical firms. Table~\ref{tab:asym} reports the resulting output coefficients and aggregate crossover scales. The heterogeneous equilibrium quantities vary across firms, but the crossover pattern remains essentially unchanged: the trapped-ion calibration crosses near $10^{12}$, whereas the Rydberg calibration requires a scale near $10^{18}$. Thus, the qualitative energy-advantage conclusion is driven by the relative scaling laws and hardware coefficients, rather than by exact within-group symmetry. In Table~\ref{tab:asym}, the output coefficient
$k_i=q_i^*(a)/a$ is computed from the general system
$q^*(a)=\Gamma^{-1}(\lambda a)$. The crossover values are obtained by solving
the aggregate quantum-classical energy equality under the baseline hardware
coefficients.

\begin{table}[tb]
\caption{Deterministic asymmetric $20$-firm scenario check.}
\label{tab:asym}
\centering
\small
\begin{tabular}{lcccc}
\toprule
Technology & $\lambda_i$ range & $\theta_i$ range & $k_i$ range & Mean $k_i$ \\
\midrule
Quantum & $[0.980,1.020]$ & $[1.963,2.040]$ & $[0.0316,0.0528]$ & $0.0435$ \\
Classical & $[0.980,1.020]$ & $[1.963,2.040]$ & $[0.0328,0.0529]$ & $0.0436$ \\
\midrule
Trapped-ion crossover & \multicolumn{4}{c}{$1.28\times 10^{12}$} \\
Rydberg crossover & \multicolumn{4}{c}{$2.77\times 10^{18}$} \\
\bottomrule
\end{tabular}
\end{table}

\subsection{Remark on Simulation Implications}
The model suggests that energy advantage should not be treated as a binary label attached to a hardware platform. It is an equilibrium outcome that depends jointly on technology coefficients, market scale, and the competitive environment. This point is particularly relevant for simulation researchers and practitioners evaluating hybrid quantum-classical workflows, including those based on variational quantum algorithms \shortcite{cerezo2021variational} and quantum computational chemistry \shortcite{mcardle2020quantum}. A platform that is not energy efficient for small instances may still become attractive when the workload scale is sufficiently large.

The framework is also useful as a building block for more elaborate simulation studies. Because the equilibrium is available in closed form or through simple boundary checks, it can be embedded inside a larger digital twin of a computing-service ecosystem, a simulation-based capacity planning model, or a stochastic demand model with time-varying energy budgets. In that sense, the contribution is not limited to the specific calibration used here; it is a reusable scenario-analysis layer for simulation and quantum computing applications.

For a concrete application mapping, consider quantum chemistry. In an active-space electronic-structure problem, $q$ can be interpreted as the effective Hilbert-space dimension induced by the active space, with $n=\log_2 q$ corresponding to the logical qubit count after mapping the Hamiltonian to a qubit representation. An analyst could estimate $\beta_q$ and $\mu$ by running a sequence of instances at a fixed accuracy tolerance and regressing measured quantum energy, computed as device power multiplied by the total circuit duration including repetitions or shots, state preparation, measurement, and readout, on the logical problem size $n$. The classical parameters $\beta_c$ and $\alpha$ could be estimated analogously from measured HPC energy consumption for a classical baseline on the same instance sequence. Under this interpretation, the crossover scale $a^\dagger$ becomes an application-specific scenario estimate for when the fitted quantum energy curve falls below the fitted classical benchmark in the competitive environment specified by the model.

Several limitations should be kept in mind. The workload index $q$ is a reduced-form proxy for computational difficulty. The model abstracts from error-correction overhead, queuing delay, capital expenditure, electricity price uncertainty, and endogenous technology adoption. It also evaluates per-firm energy in a static equilibrium, so it does not capture transient scheduling effects or hardware utilization dynamics. Accordingly, the reported crossover scales should be interpreted as scenario-analysis outputs under stated calibrations, not as forecasts of the future quantum computing market. These limitations are deliberate because the goal of this paper is to provide a tractable first model that can be extended.

\section{CONCLUSION}
This paper presented a simulation-based framework for evaluating the energy implications of quantum-classical competition. We modeled quantum and classical computing service providers as differentiated Cournot competitors subject to technology-specific energy scaling laws (polylogarithmic for quantum platforms and polynomial for classical emulation) and derived closed-form equilibrium quantities via a tractable two-group reduction. We further characterized the capacity-constrained Nash equilibrium induced by per-firm energy budgets and proved that a crossover demand scale exists beyond which quantum service consumes less energy than its classical counterpart.

Our deterministic calibration, grounded in physical parameters for trapped-ion and Rydberg hardware, showed that the crossover occurs near $10^{12}$ for ion traps and $10^{18}$ for Rydberg platforms. A $40{,}000$-scenario Monte Carlo study confirmed that these orderings are robust to substantial uncertainty in market structure, substitution parameters, and hardware coefficients, and translated the asymptotic result into an operational risk profile.

The central takeaway is that quantum energy advantage is not a binary label but a scale-dependent equilibrium outcome: the quantum constant $\beta_q$ may far exceed the classical constant $\beta_c$ at small workloads, yet the polylogarithmic growth rate eventually dominates, making quantum computing the more energy-efficient technology at sufficiently large scales. This insight is directly relevant to practitioners evaluating hybrid quantum-classical workflows and to policymakers assessing the sustainability profile of emerging computing infrastructures.

Several directions merit future investigation. Incorporating stochastic arrival processes, dynamic scheduling, and hybrid workload-routing rules would connect the model more tightly to simulation-based resource management. Extending the framework to account for error-correction overhead, endogenous technology adoption, and time-varying energy budgets would further enhance its applicability as a reusable scenario-analysis layer for digital-twin applications in quantum computing.

\appendix
\section{GENERAL ASYMMETRIC FORMULATION}
\label{app:asym}
The model can be extended to a fully asymmetric $N$-firm game. Let $q=(q_1,\ldots,q_N)^\top\in \R^N$ and define profits
\begin{equation}
\pi_i(q)=q_i\left(a_i-\sum_{j=1}^N \gamma_{ij}q_j\right),
\qquad \gamma_{ii}=\theta_i>0.
\label{eq:appendix-profit}
\end{equation}
The first-order conditions are
\begin{equation}
2\theta_i q_i+\sum_{j\neq i}\gamma_{ij}q_j=a_i,
\qquad i=1,\ldots,N.
\label{eq:appendix-foc}
\end{equation}
If we define the matrix $\Gamma\in\R^{N\times N}$ by
\begin{equation}
\Gamma_{ii}=2\theta_i,
\qquad
\Gamma_{ij}=\gamma_{ij}\quad (i\neq j),
\label{eq:appendix-Gamma}
\end{equation}
then the interior equilibrium solves
\begin{equation}
\Gamma q^*=a.
\label{eq:appendix-system}
\end{equation}
Hence $q^*=\Gamma^{-1}a$ whenever $\Gamma$ is invertible. A sufficient condition is strict diagonal dominance,
\begin{equation}
2\theta_i>\sum_{j\neq i}|\gamma_{ij}|
\qquad \text{for all } i,
\label{eq:appendix-dd}
\end{equation}
which guarantees uniqueness of the interior solution. The two-group specification used in the main text is a parsimonious special case of this more general model.

\section{TWO-GROUP BLOCK-MATRIX DERIVATION}
\label{app:block}
This appendix shows how Proposition~1 arises from the general $N$-firm model by exploiting the within-group symmetry structure.
Partition the $N=n_q+n_c$ firms into a quantum block ($i=1,\ldots,n_q$) and a classical block ($i=n_q+1,\ldots,N$). The interaction matrix takes the block form
\begin{equation}
\Gamma = \begin{pmatrix}
\Gamma_{qq} & \Gamma_{qc}\\
\Gamma_{cq} & \Gamma_{cc}
\end{pmatrix},
\label{eq:block-Gamma}
\end{equation}
where $\Gamma_{qq}\in\R^{n_q\times n_q}$ has diagonal entries $2\theta_q$ and off-diagonal entries $\gamma_{qq}$, and $\Gamma_{cc}\in\R^{n_c\times n_c}$ has diagonal entries $2\theta_c$ and off-diagonal entries $\gamma_{cc}$. The cross-blocks satisfy $\Gamma_{qc}=\gamma_{qc}J_{n_q,n_c}$ and $\Gamma_{cq}=\gamma_{cq}J_{n_c,n_q}$, where $J_{n_1,n_2}$ denotes the $n_1\times n_2$ all-ones matrix.

By the block-matrix inversion formula,
\begin{equation}
\Gamma^{-1} = \begin{pmatrix}
\Gamma_{qq}^{-1}+\Gamma_{qq}^{-1}\Gamma_{qc}S_c^{-1}\Gamma_{cq}\Gamma_{qq}^{-1}
& -\Gamma_{qq}^{-1}\Gamma_{qc}S_c^{-1}\\[4pt]
-S_c^{-1}\Gamma_{cq}\Gamma_{qq}^{-1}
& S_c^{-1}
\end{pmatrix},
\label{eq:block-inv}
\end{equation}
where $S_c=\Gamma_{cc}-\Gamma_{cq}\Gamma_{qq}^{-1}\Gamma_{qc}$ is the Schur complement.

Because $\Gamma_{qq}$ is a symmetric matrix with constant diagonal $2\theta_q$ and constant off-diagonal $\gamma_{qq}$, its inverse has the form
$(\Gamma_{qq}^{-1})_{ij}=\phi_{qq}\delta_{ij}+\psi_{qq}(1-\delta_{ij}),$
where
\begin{equation}
\phi_{qq}=\frac{2\theta_q+(n_q-2)\gamma_{qq}}{(2\theta_q-\gamma_{qq})(2\theta_q+(n_q-1)\gamma_{qq})},
\qquad
\psi_{qq}=\frac{-\gamma_{qq}}{(2\theta_q-\gamma_{qq})(2\theta_q+(n_q-1)\gamma_{qq})}.
\label{eq:phipsi-qq}
\end{equation}
The row sums of $\Gamma_{qq}^{-1}$ are
\begin{equation}
\phi_{qq}+(n_q-1)\psi_{qq}=\frac{1}{2\theta_q+(n_q-1)\gamma_{qq}}=\frac{1}{A_q},
\label{eq:rowsum}
\end{equation}
with $A_q$ defined as in \eqref{eq:ABC}. Analogously, $A_c=2\theta_c+(n_c-1)\gamma_{cc}$.

Since $\Gamma_{qc}$ and $\Gamma_{cq}$ are constant matrices, the product $\Gamma_{cq}\Gamma_{qq}^{-1}\Gamma_{qc}$ is a constant matrix with every entry equal to $\gamma_{qc}\gamma_{cq}\,n_q/A_q$. Hence the Schur complement $S_c$ is again a constant-diagonal, constant-off-diagonal matrix, allowing straightforward inversion.

At a symmetric equilibrium where $a_i=a_q$ for quantum firms and $a_i=a_c$ for classical firms, the system $\Gamma q^*=a$ reduces to two scalar equations:
\begin{equation}
A_qq_q+n_c\gamma_{qc}q_c=a_q,
\qquad
n_q\gamma_{cq}q_q+A_cq_c=a_c,
\label{eq:reduced-system}
\end{equation}
which is precisely \eqref{eq:foctwo}. Solving by Cramer's rule with $\Delta=A_qA_c-n_qn_c\gamma_{qc}\gamma_{cq}$ recovers \eqref{eq:qqstar}-\eqref{eq:qcstar}. The equilibrium quantities and profits in terms of the original variables are
\begin{align}
q_q^*&=\frac{a_q A_c - a_c n_c\gamma_{qc}}{\Delta},
\qquad
q_c^*=\frac{a_c A_q - a_q n_q\gamma_{cq}}{\Delta},
\label{eq:qstar-original}\\[4pt]
\pi_q^*&=\frac{\theta_q\bigl(a_q A_c - a_c n_c\gamma_{qc}\bigr)^2}{\Delta^2},
\qquad
\pi_c^*=\frac{\theta_c\bigl(a_c A_q - a_q n_q\gamma_{cq}\bigr)^2}{\Delta^2}.
\label{eq:pistar-original}
\end{align}
Note that $\pi_{q,c}^*>0$ whenever $\theta_{q,c}>0$, which is consistent.

In the following two special cases we set $\gamma_{qc}=\gamma_{cq}$ for brevity.

\textbf{Special case 1} ($n_q=1$, $n_c=1$). The duopoly yields
\begin{equation}
q_q^*=\frac{2a_q\theta_c-a_c\gamma_{qc}}{4\theta_c\theta_q-\gamma_{qc}^2},
\qquad
q_c^*=\frac{2a_c\theta_q-a_q\gamma_{qc}}{4\theta_c\theta_q-\gamma_{qc}^2}.
\label{eq:duopoly}
\end{equation}

\textbf{Special case 2} ($n_q=2$, $n_c=2$). The equilibrium is
\begin{equation}
q_q^*=\frac{a_q(2\theta_c+\gamma_{cc})-2a_c\gamma_{qc}}{(2\theta_c+\gamma_{cc})(2\theta_q+\gamma_{qq})-4\gamma_{qc}^2},
\qquad
q_c^*=\frac{a_c(2\theta_q+\gamma_{qq})-2a_q\gamma_{qc}}{(2\theta_c+\gamma_{cc})(2\theta_q+\gamma_{qq})-4\gamma_{qc}^2}.
\label{eq:2v2}
\end{equation}

\section{ENERGY ADVANTAGE ANALYSIS}
\label{app:supremacy}
This appendix provides the detailed analysis underlying the energy advantage claims.

\subsection{Profit Advantage Under a Common Energy Budget}
Suppose both technology classes face the same energy budget $E$. The capacity ceilings are
\begin{equation}
q_q^E=2^{(E/\beta_q)^{1/\mu}},
\qquad
q_c^E=\bigl(E/\beta_c\bigr)^{1/\alpha}.
\label{eq:qE-def}
\end{equation}
When both constraints bind ($q_q^E\ll q_q^*$ and $q_c^E\ll q_c^*$), the constrained profit functions become
\begin{align}
\pi_q^F&=a_q q_q^E - n_c\gamma_{qc}q_c^E q_q^E - \bigl(\theta_q+(n_q-1)\gamma_{qq}\bigr)(q_q^E)^2,
\label{eq:piqF}\\
\pi_c^F&=a_c q_c^E - n_q\gamma_{cq}q_c^E q_q^E - \bigl(\theta_c+(n_c-1)\gamma_{cc}\bigr)(q_c^E)^2.
\label{eq:picF}
\end{align}
For comparable market parameters ($n_q\approx n_c$, $a_q\approx a_c$), the quantum profit dominates whenever $q_q^E\gg q_c^E$. In the limit $q_c^E\to 0^+$, the condition $\pi_q^F\gg\pi_c^F$ reduces to
\begin{equation}
a_q q_q^E\gg a_c q_c^E,
\label{eq:profit-dom}
\end{equation}
which holds because the quantum capacity ceiling grows exponentially in $E^{1/\mu}$ while the classical ceiling grows only polynomially. Thus, under a common energy budget and at sufficiently large scale, the representative quantum firm captures a larger feasible output and therefore earns higher profit.

\subsection{Energy Cost at Equilibrium}
Away from the capacity-binding regime, one can compare the energy cost each technology incurs at the unconstrained Nash equilibrium:
\begin{equation}
E_q=\beta_q(\log_2 q_q^*)^{\mu},
\qquad
E_c=\beta_c(q_c^*)^{\alpha}.
\label{eq:Eqc-NE}
\end{equation}
Since $q_q^*$ and $q_c^*$ share the same denominator $\Delta$ and have comparable numerators when the market parameters satisfy $a_q\approx a_c$ and $\theta_q\approx\theta_c$, the equilibrium quantities are of the same order: $q_q^*\approx q_c^*$. Consequently,
\begin{equation}
\frac{E_q}{E_c}=\frac{\beta_q}{\beta_c}\cdot\frac{(\log_2 q_q^*)^{\mu}}{(q_c^*)^{\alpha}}
\longrightarrow 0
\qquad\text{as }a\to\infty.
\label{eq:Eratio}
\end{equation}

\subsection{Critical Scale for Energy Advantage}
In the many-firm limit ($n_q,n_c\gg 1$) with $a_q=a_c=a$ and $\gamma_{qc}\ll\gamma_{qq},\gamma_{cc}$, the equilibrium quantities simplify to
\begin{equation}
q_q^*\approx\frac{a}{\gamma_{qq}n_q},
\qquad
q_c^*\approx\frac{a}{\gamma_{cc}n_c}.
\label{eq:qstar-large}
\end{equation}
Substituting into the energy ratio and requiring $E_q<E_c$ yields
\begin{equation}
\frac{\bigl[\log_2(a/(\gamma_{qq}n_q))\bigr]^{\mu}}{\bigl(a/(\gamma_{cc}n_c)\bigr)^{\alpha}}
< \frac{\beta_c}{\beta_q}\ll 1.
\label{eq:critical-condition}
\end{equation}
Because the left-hand side decreases to zero as $a\to\infty$, there exists a critical demand scale $a^*$ satisfying
\begin{equation}
\frac{\bigl[\log_2(a^*/(\gamma_{qq}n_q))\bigr]^{\mu}}{\bigl(a^*/(\gamma_{cc}n_c)\bigr)^{\alpha}}
= \frac{\beta_c}{\beta_q}.
\label{eq:astar}
\end{equation}
For $a<a^*$ (small-scale computation), classical methods are more energy efficient; for $a>a^*$ (large-scale computation), quantum platforms achieve an energy advantage despite their larger per-unit constant $\beta_q\gg\beta_c$.

\section*{AUTHOR BIOGRAPHIES}

\noindent {\bf \MakeUppercase{Junyu Liu}} is an Assistant Professor in the Department of Computer Science at the University of Pittsburgh, Pittsburgh, PA, USA. He holds a Ph.D. in Physics from the California Institute of Technology. His research focuses on quantum computing, quantum communication, machine learning, and quantum sensing. His e-mail address is \email{junyuliu@pitt.edu} and his website is \url{https://sites.google.com/view/junyuliu/main}.\\

\noindent {\bf \MakeUppercase{Hansheng Jiang}} is an Assistant Professor of Operations Management and Statistics at the Rotman School of Management, University of Toronto, Toronto, ON, Canada. She holds a Ph.D. in Operations Research from the University of California, Berkeley. Her research explores methodological and algorithmic foundations for AI-powered and data-driven decision-making under uncertainty, with applications in sustainability, logistics, supply chains, and revenue management. Her e-mail address is \email{hansheng.jiang@rotman.utoronto.ca} and her website is \url{https://hanshengjiang.github.io/}.\\

\noindent {\bf \MakeUppercase{Zuo-Jun Max Shen}} is Chair Professor in the Department of Data and Systems Engineering and Department of Innovation and Information Management at the University of Hong Kong, Hong Kong, China, and Professor Emeritus in the Department of Industrial Engineering and Operations Research at the University of California, Berkeley. His research interests include logistics and supply chain management, data-driven decision-making, and energy systems optimization. He is a Fellow of INFORMS and POMS. His e-mail address is \email{maxshen@hku.hk} and his website is \url{https://zj-maxshen.github.io/}.\\

\end{document}